\documentclass[useAMS,usenatbib]{mn2e}
\usepackage{scicite}
\onecolumn
\title[]{Self-Similar Solutions of Viscous-Resistive
 ADAFs With Poloidal Magnetic Fields
 }

\author[J. Ghanbari , F. Salehi and S. Abbassi]{J.
Ghanbari$^{1,2,3}$\thanks{E-mail: ghanbari@ferdowsi.um.ac.ir}
, F. Salehi$^{3}$\thanks{E-mail:fsalehi@wali.um.ac.ir} and S. Abbassi$^{4}$\thanks{E-mail:sabbassi@dubs.ac.ir}\\
$^{1}$Department of Physics, School of Sciences, Ferdowsi University of Mashhad, Mashhad, 91775-1436, Iran\\
$^{2}$Department of Physics and Astronomy, San Francisco State
University, 1600 Holloway, San Francisco, CA 94132\\
$^{3}$Department of Physics, Khayyam Institute of High Education,
Mashhad, Iran\\
$^4$Department of Physics ,Damghan University of Basic Sciences,
Damghan, Iran}
\begin{document}
\date{}

\pagerange{\pageref{firstpage}--\pageref{lastpage}} \pubyear{2007}

\maketitle \label{firstpage}

\begin{abstract}
We carry out the self-similar solutions of viscous-resistive
accretion flows around a magnetized compact object. We consider an
axisymmetric, rotating, isotheral steady accretion flow which
contains a poloidal magnetic field of the central star. The dominant
mechanism of energy dissipation is assumed to be the turbulence
viscosity and magnetic diffusivity due to magnetic field of the
central star. We explore the effect of viscosity on a rotating disk
in the presence of constant magnetic diffusivity. We show that the
dynamical quantities of ADAFs are sensitive to the advection and
viscosity parameters. Increase of the $\alpha$ coefficient in the
$\alpha$-prescription model decreases the radial velocity and
increases the density of the flow. It also affects the poloidal
magnetic field considerably.
\end{abstract}

\section{INTRODUCTION}
Accretion onto black holes has been intensely studied for the last
three decades (see Kato et al. 1998 for a review), and several types
of models were proposed. Standard accretion disk model of Shakura
$\&$ Sunyaev (1973) has been very useful in interpretation of
observations in binary systems and active galactic nuclei, AGN
(Pringle 1981), is based on a number of simplifying assumptions. In
particular, the flow is assumed to be geometrically thin and with a
Keplerian angular velocity distribution. This assumption allows
gradient terms in the differential equations describing the flow to
be neglected, reducing them to a set of algebraic equations and
thereby fixes the angular momentum distribution of the flow. For low
accretion rates, $\dot{M}$, this assumption is generally considered
to be reasonable. Since the end of seventies, however, it has been
realized that for high accretion rates, advection of energy with the
flow can crucially modifies the properties of the innermost parts of
accretion disks around black holes.

The natural improvement over the Shakura-Sunyeav model was to
consider the case when cooling is less efficient than viscous
heating. This may happen in two cases: either when the disk is
extremely optically thick and the radiation is trapped for a
timescale longer than the accretion timescale( see Abramowicz  et
al. 1988) or when it is extremely optically thin, a regime in which
cooling processes are inefficient ( Narayan $\&$ Yi 1995b).

The study of advection accretion flows around low-luminosity black
hole candidates and neutron stars are currently a very active field
of research, both theoretically and observationally (see Narayan et
al. 1998 for a review). Observational evidences for the existence of
low luminosity black holes at the center of galaxies and in the
active galactic nuclei AGN (Cherepashchuk 1996; Ho 1999) make
necessary to revise theoretical models of the accretion disks.
Thereby development of the subject of advection-dominated accretion
flow (ADAF) in recent years lead to global solutions of advection
accretion disks around accreting black hole systems and neutron
stars. In this case, viscously generated internal energy is not
radiated away efficiently as the gas falls into the potential well
of the central mass (as in the standard thin disks model, Shakura \&
Sunyaev 1973) but retained within the accreting gas and advected
radially inward (Narayan \& Yi 1994, hereafter NY1994) and might
eventually be lost into the central object or in contrast, a
considerable portion of it might give rise to wind onto black holes
and neutron stars (Blandford \& Begelman 1999). By definition,
Advection-Dominated Accretion Disks, ADAFs, have very low radiative
efficiency as a consequence they can be considerably hotter than the
gas flow in the standard thin disk models (Narayan \& Yi 1995 a,b)
and therefore they are ultra-dim for their accretion rates (Phinney
1981; Rees et al. 1982). On the other hand, since all the internal
energy is stored as thermal energy, the gas becomes extremely hot
and the kinetic temperature of ions approaches the virial limit what
means that even at high initial angular momentum, the disk becomes
very thick, forming practically a quasi-spherical accretion flow
(Narayan \& Yi 1995a). A general description of an advective
accretion flow around a compact star was put forward by Narayan \&
Yi (1994, 1995a): they parameterized the degree of advection with
one parameter $f$, defined as the ratio between the thermal energy
stored in the disk and advected the central object ( not radiated),
and the total thermal energy generated by viscosity at each radius.
The general result obtained both from self-similar solutions and
numerical calculations is that high advection $(f\sim1)$ produces a
hotter, thicker disk with a larger infall radial velocity and
conversely a sub-Keplerian circular velocity.

A remarkable problem arises when the accretion disk was threaded by
magnetic field. There are good reasons for believing that the
magnetic fields are important in accretion processes in ADAFs. Some
authors tried to study magnetized accretion flows analytically. For
example ,Kaburaki (2000) presented a set of analytical solutions for
a fully advective accretion flow in a global magnetic field.
Shadmehri (2004), hereafter SH2004, extended this analysis by
considering a non-constant resistivity. He obtained a set of
self-similar solutions in spherical coordinates that described
quasi-spherical magnetized accretion flow. The full account of the
processes, connected with a presence of magnetic field in the flow,
is changing considerably the picture of the accretion flow.

In ADAF models, energy dissipation in the accretion flow can be
assumed to be due to turbulent viscosity and electrical resistivity.
Under some conditions, it is important that we consider the effect
of resistivity on accretion flows. Kuwabara et al. (2000) showed the
results of global MHD simulations of an accretion flow initially
threaded by large-scale poloidal magnetic fields including the
effects of magnetic turbulent diffusivity. They found the importance
of strength of magnetic diffusivity when they studied it in
magnetically driven mass accretion. They pointed out that the mass
outflow depends on the strength of magnetic diffusivity, so that for
a highly diffusive disk, no outflow takes place. Thereby in this
paper, we want to explore how the structure of a steady state thick
disk depends on its resistivity and viscosity so we pursue SH2004's
work, resistive disks, when accretion flow experiences the rotation
as well as viscosity dissipation. We consider ADAFs with the pure
inflow and investigate the effect of viscosity on some physical
quantities of the flows such as the radial and angular velocities,
the density and the magnetic field flux. In order to study the
dynamics of these flows, several simplified assumptions must be made
in the analysis. The fluid is treated at least approximately as
non-relativistic and also a poloidal model is adopted for
electromagnetic field in which, it has a poloidal component in the
disk. However, we will present self-similar solutions for
viscous-resistive ADAFs.

This paper is organized as follows. Section 2, we present the
equations of magnetohydrodynamics as the basic equations. General
principles are presented in section 3. We show that the equations
can be solved using the self-similar method and the numerical
solutions are discussed in section 4 followed by results in section
5, summery and conclusion in section 6.
\section{The Basic Equations}
As we stated in introduction, we are interested in constructing a
model for describing accretion disks in global magnetic fields.
The macroscopic behavior of such flows can be studied by MHD
equations. For simplicity the self-gravity of the disks and the
effect of general relativity have been neglected. The flow is
described in terms of the flow-frame time derivative or co-moving
derivative, i.e. $\frac{D}{Dt}$ that defined as:
$\frac{D}{Dt}=\frac{\partial }{\partial
t}+(\textbf{u}\cdot\nabla)$. So, we can describe the accretion
flows by the fundamental governing equations which are written by
the equation of continuity:
\begin{equation}
\frac{D\rho}{Dt}+\rho{\nabla}\cdot\textbf{u}=0\label{con1},
\end{equation}

the equation of motion:
\begin{eqnarray}
\rho\frac{D\textbf{u}}{Dt}=-{\nabla}p-\rho{\nabla}\Phi
+\mu\nabla^2\textbf{u}+(\mu_b+\frac{1}{3}){\nabla}({\nabla}
\cdot\textbf{u})+\frac{1}{4\pi}\textbf{J}\times\textbf{B}\label{mot},
\end{eqnarray}

the equation of energy:
\begin{equation}
\rho\left[\frac{D\epsilon}{Dt}+p\frac{D}{Dt}(\frac{1}{\rho})\right]=
Q^{+}-Q^{-}=Q^{adv}\label{ene1},
\end{equation}

and the Maxwell`s equations:
\begin{equation}
{\nabla}\cdot\textbf{B}=0\label{max1},
\end{equation}
\begin{equation}
\frac{D\textbf{B}}{Dt}={\nabla}\times(\textbf{u}\times\textbf{B})+
\eta\nabla^2\textbf{B}\label{max2},
\end{equation}
where $\rho$ is the density of the gas, p the gas pressure,
$\epsilon$ the internal energy, \textbf{u} the flow velocity,
\textbf{B} the magnetic field, $\textbf{J}={\nabla}\times\textbf{B}$
the current density, $\eta$ the magnetic diffusivity in which for
simplicity it is assumed to be a constant parameter (see, e.g.,
Kaburaki 2000), $\mu$ and $\mu_b$ are the shear and bulk
viscosities, $Q^{adv}$ represents the advective transport of energy
and is defined as the difference between the viscous heating rate
$Q^+$ and radiative cooling rate $Q^-$. We neglect self-gravity so
that $\Phi$ is assumed to be due to a central object. Also we
neglect radiation pressure in the equations because in the optically
thin ADAFs, $P^{gas}\gg P^{rad}$.

We employ the parameter $f=\frac{Q^{adv}}{Q^+}$(e.g., NY1994) to
measure the degree to which the accretion flow is
advection-dominated. In general, it will vary with r and depends on
the details of the heating and cooling processes. For simplicity, it
is assumed a constant. So, for advection-dominated flows we have
$Q^{adv}\simeq Q^{+}\gg Q^{-}$. This corresponds to an optically
thin ADAF where the viscous energy is stored in the gas as internal
energy and the amount of cooling is negligible compared to the
heating. In this case, the accreting gas has a very low density and
accretion rates are low, sub-Eddington (Ichimaru 1977; Rees et al.
1982; NY1994; Abramowics et al. 1995). Correspondingly, for
radiative-cooling-dominated flows we have $Q^{+}\simeq Q^{-}\gg
Q^{adv}$. This corresponds to a cooling-dominated flow where the
viscous energy is released in the gas as radiative energy and the
amount of energy advected is negligible. The optically thick
Shakura-Sunyaev disks correspond to this case.

Now, we formulate the basic equations (\ref{con1})-(\ref{max2}) in
spherical polar coordinates as follows:
\begin{equation}
\frac{\partial\rho}{\partial
t}+\frac{1}{r^2}\frac{\partial}{\partial r}(r^2\rho
u_r)+\frac{1}{r}\frac{\partial}{\partial\theta}(\rho
u_\theta)=0\label{con2},
\end{equation}

The three components of the momentum equations give (e.g., Mihalas
$\&$ Mihalas 1984):
\begin{eqnarray}
\lefteqn{\rho\left[\frac{\partial u_r}{\partial t}+u_r\frac{\partial
u_r}{\partial r}+\frac{u_\theta}{r}\frac{\partial
u_r}{\partial\theta}+\frac{u_\phi}{r\sin\theta}\frac{\partial
u_r}{\partial\phi}-\frac{1}{r}(u_\theta^2+u_\phi^2)\right]
=-\frac{GM\rho}{r^2}-\frac{\partial p}{\partial
r}}\nonumber\\
&&+\frac{\partial}{\partial r}\left[2\mu\frac{\partial u_r}{\partial
r}+(\mu_b-\frac{2}{3}\mu)
\left(\frac{1}{r^2}\frac{\partial}{\partial
r}(r^2u_r)+\frac{1}{r\sin\theta}\frac{\partial}{\partial\theta}(u_\theta\sin\theta)
+\frac{1}{r\sin\theta}\frac{\partial u_\phi}{\partial\phi}\right)\right]\nonumber\\
&&+\frac{1}{r}\frac{\partial}{\partial\theta}\left[r\mu\frac{\partial}{\partial
r}\left(\frac{u_\theta}{r}\right)+\frac{\mu}{r}\frac{\partial
u_r}{\partial\theta}\right]+\frac{1}{r\sin\theta}\frac{\partial}{\partial\phi}
\left[\frac{\mu}{r\sin\theta}\frac{\partial u_r}{\partial\phi}
+r\mu\frac{\partial}{\partial
r}\left(\frac{u_\phi}{r}\right)\right]\nonumber\\
&&+\frac{\mu}{r}\left[4r\frac{\partial}{\partial
r}\left(\frac{u_r}{r}\right)-\frac{2}{r\sin\theta}\frac{\partial}{\partial\theta}(u_\theta\sin\theta)-
\frac{2}{r\sin\theta}\ \frac{\partial
u_\phi}{\partial\phi}+r\cot\theta\frac{\partial}{\partial
r}\left(\frac{u_\theta}{r}\right)+\frac{\cot\theta}{r}\frac{\partial
u_r}{\partial\theta}\right]\nonumber\\
&&+\frac{1}{4\pi}\left\{-B_\phi\frac{1}{r}\frac{\partial}{\partial
r}(rB_\phi)-B_\theta\frac{1}{r}\left[\frac{\partial}{\partial
r}(rB_\theta)-\frac{\partial}{\partial\theta}B_r\right]\right\}\label{motr}
\end{eqnarray}

\begin{eqnarray}
\lefteqn{\rho\left[\frac{\partial u_\theta}{\partial
t}+u_r\frac{\partial u_\theta}{\partial
r}+\frac{u_\theta}{r}\frac{\partial
u_\theta}{\partial\theta}+\frac{u_\phi}{r\sin\theta}\frac{\partial
u_\theta}{\partial\phi}+\frac{1}{r}(u_ru_\theta-u_\phi^2\cot\theta)\right]}\nonumber\\
&&=-\frac{1}{r}\frac{\partial
p}{\partial\theta}+\frac{1}{r}\frac{\partial}{\partial
r}\left[r\mu\frac{\partial}{\partial
r}\left(\frac{u_\theta}{r}\right)+\frac{\mu}{r}\frac{\partial
u_r}{\partial\theta}\right]\nonumber\\
&&+\frac{1}{r}\frac{\partial}{\partial
\theta}\left[\frac{2\mu}{r}(\frac{\partial
u_\theta}{\partial\theta}+(\mu_b-\frac{2}{3}\mu)
\left(\frac{1}{r^2}\frac{\partial}{\partial
r}(r^2u_r)+\frac{1}{r\sin\theta}\frac{\partial}{\partial\theta}(u_\theta\sin\theta)
+\frac{1}{r\sin\theta}\frac{\partial u_\phi}{\partial\phi}\right)\right]\nonumber\\
&&+\frac{1}{r\sin\theta}\frac{\partial}{\partial\phi}
\left[\frac{\mu\sin\theta}{r}\frac{\partial}{\partial\theta}
\left(\frac{u_\phi}{\sin\theta}\right)+\frac{\mu}{r\sin\theta}\frac{\partial
u_\theta}{\partial\phi}\right]\nonumber\\
&&+\frac{\mu}{r}\left\{\frac{2\cot\theta}{r}
\left[\sin\theta\frac{\partial}{\partial\theta}
\left(\frac{u_\theta}{\sin\theta}\right)-
\frac{1}{\sin\theta}\frac{\partial
u_\phi}{\partial\phi}\right]+3r\frac{\partial}{\partial
r}\left(\frac{u_\theta}{r}\right)+\frac{3}{r}\frac{\partial
u_r}{\partial\theta}\right\}\nonumber\\
&&+\frac{1}{4\pi}\left\{B_r\frac{1}{r}\left[\frac{\partial}{\partial
r}(rB_\theta)-\frac{\partial
B_r}{\partial\theta}\right]-B_\phi\frac{1}{r\sin\theta}\frac{\partial}{\partial
\theta}(B_\phi\sin\theta)\right\}\label{mottheta}
\end{eqnarray}

\begin{eqnarray}
\lefteqn{\rho\left[\frac{\partial u_\phi}{\partial
t}+u_r\frac{\partial u_\phi}{\partial
r}+\frac{u_\theta}{r}\frac{\partial
u_\phi}{\partial\theta}+\frac{u_\phi}{r\sin\theta}\frac{\partial
u_\phi}{\partial\phi}+\frac{u_\phi}{r}(u_r+u_\theta\cot\theta)\right]
=-\frac{1}{r\sin\theta}\frac{\partial
p}{\partial\phi}}\nonumber\\
&&+\frac{\partial}{\partial
r}\left[\frac{\mu}{r\sin\theta}\frac{\partial u_r}{\partial
\phi}+r\mu\frac{\partial }{\partial
r}\left(\frac{u_\phi}{r}\right)\right]+\frac{1}{r}\frac{\partial}{\partial
\theta}\left[\frac{\mu\sin\theta}{r}\frac{\partial}{\partial
\theta}\left(\frac{u_\phi}{\sin\theta}\right)+r\mu\frac{\partial
}{\partial
r}\left(\frac{u_\phi}{r}\right)\right]\nonumber\\
&&+\frac{1}{r\sin\theta}\frac{\partial}{\partial
\phi}\left[\frac{2\mu}{r}\left(\frac{1}{\sin\theta}\frac{\partial
u_\phi}{\partial\phi}+u_r+u_\theta\cot\theta\right)+(\mu_b-\frac{2}{3}\mu)\right.\nonumber\\
&&\left.\times\left(\frac{1}{r^2}\frac{\partial}{\partial
r}(r^2u_r)+\frac{1}{r\sin\theta}\frac{\partial}{\partial\theta}(u_\theta\sin\theta)
+\frac{1}{r\sin\theta}\frac{\partial u_\phi}{\partial\phi}\right)\right]\nonumber\\
&&+\frac{1}{r\sin\theta}\frac{\partial}{\partial\phi}
\left[\frac{\mu\sin\theta}{r}\frac{\partial}{\partial\theta}
\left(\frac{u_\phi}{\sin\theta}\right)+\frac{\mu}{r\sin\theta}\frac{\partial
u_\theta}{\partial\phi}\right]\nonumber\\
&&+\frac{\mu}{r}\left\{2\cot\theta
\left[\frac{\sin\theta}{r}\frac{\partial}{\partial\theta}
\left(\frac{u_\phi}{\sin\theta}\right)+
\frac{1}{\sin\theta}\frac{\partial
u_\theta}{\partial\phi}\right]+3r\frac{\partial}{\partial
r}\left(\frac{u_\phi}{r}\right)+\frac{3}{r\sin\theta}\frac{\partial
u_r}{\partial\phi}\right\}\nonumber\\
&&+\frac{1}{4\pi}\left\{B_r\frac{1}{r}\frac{\partial}{\partial
r}(rB_\phi)+B_\theta\frac{1}{r\sin\theta}\frac{\partial}{\partial
\theta}(B_\phi\sin\theta)\right\}\label{motphi}
\end{eqnarray}

while the equation of the energy gives:
\begin{eqnarray}
\lefteqn{\rho\left[\frac{\partial\epsilon}{\partial
t}+u_r\frac{\partial\epsilon}{\partial
r}+\frac{u_\theta}{r}\frac{\partial\epsilon}{\partial\theta}
+\frac{u_\phi}{r\sin\theta}\frac{\partial\epsilon}{\partial\phi}
-\frac{p}{\rho^2}\left(\frac{\partial}{\partial
t}+u_r\frac{\partial}{\partial
r}+\frac{u_\theta}{r}\frac{\partial}{\partial\theta}+\frac{u_\phi}{r\sin\theta}
\frac{\partial}{\partial\phi}\right)\rho\right]}\nonumber\\
&&=2\mu f\left\{\left(\frac{\partial u_r}{\partial
r}\right)^2+\frac{1}{r^2}\left(\frac{\partial
u_\theta}{\partial\theta}+u_r\right)^2+\left[\frac{1}{r\sin\theta}
\frac{\partial
u_\phi}{\partial\phi}+\frac{1}{r}(u_r+u_\theta\cot\theta)\right]^2\right.\nonumber\\
&&\left.+\frac{1}{2}\left[r\frac{\partial}{\partial
r}\left(\frac{u_\theta}{r}\right)+\frac{1}{r}\frac{\partial
u_r}{\partial\theta}\right]^2
+\frac{1}{2}\left[r\frac{\partial}{\partial
\phi}\left(\frac{u_\phi}{r}\right)+\frac{1}{r\sin\theta}\frac{\partial
u_r}{\partial\phi}\right]^2\right.\nonumber\\
&&\left.+\frac{1}{2}\left[\frac{\sin\theta}{r}\frac{\partial}{\partial
\theta}\left(\frac{u_\phi}{\sin\theta}\right)+\frac{1}{r\sin\theta}\frac{\partial
u_\theta}{\partial\phi}\right]^2\right\}\nonumber\\
&&+f(\mu_b-\frac{2}{3}\mu)\left(\frac{1}{r^2}\frac{\partial}{\partial
r}(r^2u_r)+\frac{1}{r\sin\theta}\frac{\partial}{\partial\theta}(u_\theta\sin\theta)
+\frac{1}{r\sin\theta}\frac{\partial
u_\phi}{\partial\phi}\right)\label{ene2}
\end{eqnarray}
also, the r-component of the induction equation
\begin{equation}
\frac{\partial B}{\partial
t}=\frac{\partial}{\partial\theta}\left\{r \sin\theta\left[u_{\rm
r}B_{\rm\theta} -\eta\frac{1}{r}\left(\frac{\partial}{\partial
r}(rB_\theta)-\frac{\partial
B_r}{\partial\theta}\right)\right]\right\},\label{Br}
\end{equation}
the $\theta$-component
\begin{equation}
\frac{\partial B}{\partial t}=\frac{\partial}{\partial r}\left\{r
\sin\theta\left[u_{\rm r}B_{\rm\theta}
-\eta\frac{1}{r}\left(\frac{\partial}{\partial
r}(rB_\theta)-\frac{\partial
B_r}{\partial\theta}\right)\right]\right\}\label{Btheta},
\end{equation}
the $\phi$-component
\begin{equation}
\frac{\partial B}{\partial t}=\frac{\partial}{\partial r}(ru_{\rm
\varphi}B_{\rm r}-ru_{\rm
r}B_{\rm\varphi})+\frac{\partial}{\partial\theta}(u_{\rm\varphi}B_{\rm\theta})
+\frac{\partial}{\partial
r}\left[\frac{\eta}{r}\frac{\partial}{\partial r}(rB_\phi)\right]
+\frac{\partial}{\partial\theta}\left[\frac{\eta}{r\sin\theta}
\frac{\partial}{\partial\theta}(B_\phi\sin\theta)\right].\label{Bphi}
\end{equation}
\section{General Principles}
As stated, we used a total of eight partial differential equations
governing the non-self gravitating flow. These equations relate 15
dependent variables: $p,\rho,\epsilon,\mu,\mu_b,\eta$ and the
components of $\textbf{u}, \textbf{J}$ and $\textbf{B}$. The
equations must be closed by specifying suitable prescriptions for
the viscosity and for the thermodynamics. Thus for the set of
equations adopted, we make the following standard assumptions:
we consider a steady ($\frac{\partial}{\partial t}=0$), rotating,
axisymmetric ($\frac{\partial}{\partial\varphi}=0$), quasi spherical
accreting flow (so it is convenient that we use spherical
coordinates ($r,\theta,\phi$) in our discussion) with Keplerian
angular velocity $\Omega_k(r)=(\frac{GM}{r^3})^{\frac{1}{2}}$ around
a central object and with a purely poloidal magnetic field threading
the disk. We assume that the fluid can be treated at least
approximately as non-relativistic.

The kinematic viscosity coefficient, $\nu=\frac{\mu}{\rho}$, is
generally parameterized using the $\alpha$-prescription
(Shakura-Sunyaev 1973),
\begin{equation}
\nu=\alpha c_sH,\label{nu1}
\end{equation}
where $H=\frac{c_s}{\Omega_k}$ is known as the vertical scale height
, $c_s=\sqrt{\frac{p}{\rho}}$ is the isothermal sound speed and the
dimensionless coefficient $\alpha$ is assumed to be independent of
r. Also it is important that we consider the effect of resistivity
on accretion flows. So, we introduce the parameter $\eta$ as the
magnetic diffusivity and insert it as a constant parameter in our
equations. Both the kinematic viscosity coefficient $\nu$ and the
magnetic diffusivity $\eta$ have the same units and are assumed to
be due to turbulence in the accretion flow. Thus it is physically
reasonable to express $\eta$ such as $\nu$ via the
$\alpha$-prescription of Shakura-Sunyaev (1973) as follows
(Bisnovatyi-Kogan \& Ruzmaikin 1976),
\begin{equation}
\eta=\eta_\circ c_sH.\label{eta1}
\end{equation}
where the dimensionless coefficient $\eta_\circ$ is assumed
independent of r. Bisnovatyi-Kogan and Ruzmaikin (1976) introduced
$\eta_\circ$ and proposed that $\eta_\circ\sim\alpha$. Substituting
$H$ and $c_s$ into the relations (\ref{nu1}) and (\ref{eta1}), we
find them locally proportional to the pressure:
\begin{equation}
\nu=\alpha\frac{p}{\rho\Omega_k}\label{nu2},
\end{equation}
\begin{equation}
\eta=\eta_\circ\frac{p}{\rho\Omega_k}\label{eta2}.
\end{equation}
Their ratio is one definition of the magnetic Prandtl number as
follows,
\begin{equation}
P_m=\frac{\nu}{\eta}=\frac{u_T\ell}{\eta}.
\end{equation}
where $u_T$ is the random velocity of diffusing particle and $\ell$
is its mean free path. For a fully ionized hydrogen plasma Prandtl
number is very large compared with unity due to the large length
scale $\ell$ of the disk. In the following we consider conditions
where $Pm\leq 1$. That means, viscous and resistive forces can be
contributed in the energy dissipation similarly when it is equal to
unity and resistive forces are dominate when $P_m$ is smaller than
unity.

To determine thermodynamical properties of the flow in the energy
equation (\ref{ene2}), we require a constitutive relation as a
function of two state variables. Therefore we choose an equation for
the internal energy as $\epsilon=\frac{p}{\rho(\Gamma-1)}$ where
$\Gamma$ is the ratio of specific heats of the gas.

To satisfy $\nabla\cdot\textbf{B}=0$, we may introduce a convenient
functional form for the magnetic field. Owning to the axisymmetry,
the magnetic field can be written as
\begin{equation}
\textbf{B}=\textbf{B}_p(r,\theta)+B_\phi(r,\theta)\textbf{e}_\phi
\end{equation}

The effect of magnetic diffusivity on magnetically driven mass
accretion was studied by Kuwabara et al. (2000). They showed that
the effects of resistivity are that magnetic field lines do not
rotate with the same angular speed as the disk matter and thus it
suppresses the injection of magnetic helicity and
magneto-centrifugal acceleration. So, neglecting the toroidal
component of field, $B_\phi$, we can express the poloidal component,
$\textbf{B}_p$, in terms of a magnetic flux function
$\Psi(r,\theta)$:
\begin{equation}
\textbf{B}=\textbf{B}_p(r,\theta)=\frac{1}{2\pi}\nabla\times
\left(\frac{\Psi}{r\sin\theta}\textbf{e}_\phi\right),
\end{equation}
in which satisfies
$\nabla\cdot\textbf{B}=\nabla\cdot\textbf{B}_p=0$. Magnetic flux
function $\Psi(r,\theta)$ is related to the magnetic vector
potential by $\Psi=rA_\phi$ with $A_\phi$ the toroidal component of
the vector potential. The magnetic flux contained inside the circle
$\rm r$=constant, $\theta=$constant is,
\begin{equation}
\int_{0}^r B_\theta(r^\prime,\theta)2\pi r^\prime
dr^\prime=2\pi\Psi(r,\theta) ~~~~(+constant),
\end{equation}
Since $B_p\cdot\nabla\Psi=0$, $\Psi$ labels magnetic lines or their
surfaces of revolution, magnetic surfaces.

Similarity, one can write the flow velocity in the form
\begin{equation}
\textbf{u}=\textbf{u}_p(r,\theta)+u_\phi(r,\theta)\textbf{e}_\phi,
\end{equation}
subsequently, we show that the poloidal component of velocity has
only a radial component $u_p=u_r$. We take $u_r$ to be a negative,
since we want to consider infall of material.

It is clear that the basic equations are nonlinear and we cannot
solve them analytically. Therefore, it is useful to have a simple
means to investigate the properties of solutions. This is most
easily done in terms of a set of dimensionless parameters which can
be expected to be similar at all times. Here, one can employ the
method of self-similar to fluid equations.

\section{Self-Similar Solutions}
To better understanding the the physical processes of our
viscous-resistive ADAF accretion disks, we seek self-similar
solutions of the above equations. The self-similar method is
familiar from its wide applications to the full set of MHD
equations. As long as we are not interested in boundaries of the
problem, such solutions can accurately describe the behavior of the
solutions in an intermediate region far from the radial boundaries.

Writing the equations in non-dimensional forms, that is, scaling all
the physical variables by their typical values, brings out the
non-dimensional variables. We can simply show that a solution of the
following forms, satisfy the equations of our model:
\begin{equation}
\rho(r,\theta)=\rho_\circ\rho(\theta)(r/r_\circ)^{-3/2},
\end{equation}
\begin{equation}
\ p(r,\theta)=p_\circ P(\theta)(r/r_\circ)^{-5/2},
\end{equation}
\begin{equation}
\ u_{\rm r}(r,\theta)= r\Omega_{\rm K}(r) U(\theta),
\end{equation}
\begin{equation}
\ u_{\rm \varphi}(r,\theta)=r\sin\theta\Omega_{\rm
K}(r)\Omega(\theta),
\end{equation}
\begin{equation}
\ B_{\rm
r}(r,\theta)=\frac{B_\circ}{2\pi\sin\theta}\frac{d\Psi(\theta)}{d\theta}(r/r_\circ)^{-5/4},
\end{equation}
\begin{equation}
\ B_{\rm
\theta}(r,\theta)=-\frac{3B_\circ\Psi(\theta)}{8\pi\sin\theta}(r/r_\circ)^{-5/4}.
\end{equation}%

where $\rho_\circ$, $p_\circ$, $B_\circ$ and $r_\circ$ provide a
non-dimensional form for our equations. Also $u_\theta=0$ is
considered, since the most important assumption we made is that
the flow is steady ($\frac{\partial}{\partial t}=0$) and also with
respect to above solutions we find $r^2\rho u_r$ is independent of
r  but it is a function of $\theta$). It represents mass
accretion rate in a given $\theta$. If we integrate this over the
angle $\theta$, we obtain the net mass accretion rate
\begin{equation}
\dot{M}=-2{\dot{M}}_\circ\int\rho U\sin\theta d\theta\label{M}
\end{equation}
where ${\dot{M}}_\circ=2\pi\rho_\circ
r^2_\circ\sqrt{\frac{GM}{r_\circ}}$.

We adopt $\alpha$-prescription (\ref{nu2}) so that
$\mu=\nu\rho=\alpha\frac{p}{\Omega_k}$. The bulk viscosity is not
usually discussed in the context of accretion flows. Thus we assume
$\mu_b=0$. Substituting the above solutions in the equations
(\ref{motr})-(\ref{Bphi}), we obtain a set of coupled ordinary
differential equations in terms of $\theta$ for the three components
of the equation of motion:
\begin{eqnarray}
\rho\left(1-\frac{U^{2}}{2}-\Omega^{2}\sin^2\theta\right)= &&C_{\rm
1}P\left(2.5-\alpha U +
\alpha\frac{dU}{d\theta}\cot\theta\right)\nonumber\\
&&+\alpha
C_{\rm 1}\frac{d}{d\theta}\left(P\frac{dU}{d\theta}\right)\nonumber\\
&&+\frac{3C_{\rm 1}\Psi}{64\pi^{3}\sin\theta}
\left[\frac{3\Psi}{16\sin\theta}-\frac{d}{d\theta}
\left(\frac{1}{\sin\theta}\frac{d\Psi}{d\theta}\right)\right],
\end{eqnarray}
\begin{eqnarray}
-\rho\Omega^{2}\sin\theta\cos\theta=&&-C_{\rm
1}\frac{dP}{d\theta}+\frac{\alpha}{2}C_{\rm 1}P
\frac{dU}{d\theta}+ \alpha C_{\rm 1}\frac{d}{d\theta}(PU)\nonumber\\
&&+\frac{C_{\rm 2}}{16\pi^{3}\sin\theta}\frac{d\Psi}{d\theta}
\left[\frac{3\Psi}{16\sin\theta}-\frac{d}{d\theta}
\left(\frac{1}{\sin\theta}\frac{d\Psi}{d\theta}\right)\right],\nonumber\\
\end{eqnarray}
\begin{eqnarray}
\frac{1}{2}\rho U\Omega\sin\theta=&&-\frac{3}{4}\alpha C_{\rm
1}P\Omega\sin\theta+\alpha C_{\rm
1}\frac{d}{d\theta}\left(P\sin\theta\frac{d\Omega}{d\theta}\right)\nonumber\\
&&+2\alpha C_{\rm 1}P\frac{d\Omega}{d\theta}\cos\theta,
\end{eqnarray}
the energy equation becomes:
\begin{eqnarray}
\rho U\frac{3\Gamma-5}{2(\Gamma-1)}=&&\alpha\rho
f\left[3U^2+\left(\frac{dU}{d\theta}\right)^2+\frac{9}{4}\Omega^2\sin\theta^2+
\left(\frac{d\Omega}{d\theta}\right)^2\sin^2\theta\right]\nonumber\\
&&+\frac{f\eta_{\rm 0}C_{\rm
2}}{16\pi^3}\left[\frac{3\Psi}{16\sin\theta}-\frac{d}{d\theta}
\left(\frac{1}{\sin\theta}\frac{d\Psi}{d\theta}\right)\right]^{2},
\end{eqnarray}
equations (\ref{Br}) and (\ref{Btheta}) give:
\begin{equation}
-\rho\frac{d}{d\theta}(\Omega\Psi)-\rho\Omega\frac{d\Psi}{d\theta}=0,
\end{equation}
the equation (\ref{Bphi}) gives:
\begin{equation}
-\frac{3\rho U\Psi}{4\eta_{\rm 0}C_{\rm
1}P\sin\theta}=\frac{3\Psi}{16\sin\theta}-\frac{d}{d\theta}
\left(\frac{1}{sin\theta}\frac{d\Psi}{d\theta}\right).
\end{equation}
where $C_{\rm
1}=\frac{p_\circ}{\rho_\circ}\left(\frac{GM}{r_\circ}\right)^{-1}=\frac{2p_\circ}{\rho_\circ
u^2_{ff}}$ and $C_{\rm
2}=\frac{B^2_\circ}{\rho_\circ}\left(\frac{GM}{r_\circ}\right)^{-1}$.

The above equations constitute a set of ordinary differential
equations for functions $U(\theta)$, $\Psi(\theta)$, $P(\theta)$ and
$\Omega(\theta)$ as follows:

\begin{eqnarray}
\frac{d^2U}{d\theta^2}=&&-\frac{2.5}{\alpha}+U-\frac{dU}{d\theta}\cot\theta-
\frac{1}{P}\frac{dP}{d\theta}\frac{dU}{d\theta}\nonumber\\
&&+ \frac{\rho}{\alpha C_{\rm 1}P}(1-\frac{U^2}{2}-\Omega^2
\sin^2\theta)+\frac{2U\rho}{\beta_\circ\eta_{\circ}\alpha C_{\rm
1}}\left(\frac{3\Psi}{8\pi P\sin\theta}\right)^2,\label{U}
\end{eqnarray}

\begin{eqnarray}
\frac{d^2\Psi}{d\theta^2}=\frac{d\Psi}{d\theta}\cot\theta+
\left(\frac{3}{16}+\frac{3U\rho}{4\eta_\circ C_{\rm
1}P}\right)\Psi,\label{Psi}
\end{eqnarray}
\begin{eqnarray}
\frac{d^2\Omega}{d\theta^2}=\frac{\rho U\Omega}{2\alpha C_{\rm
1}P}+\frac{3}{4}\Omega-3\frac{d\Omega}{d\theta}cot\theta
-\frac{1}{P}\frac{dP}{d\theta}\frac{d\Omega}{d\theta},\label{Omega1}
\end{eqnarray}
\begin{eqnarray}
\frac{dP}{d\theta}=\frac{3\alpha P}{2(1-\alpha
U)}\frac{dU}{d\theta}-\frac{3\rho U\Psi}{8\pi^2\beta_\circ\eta_\circ
C_{\rm 1}P\sin^2\theta(1-\alpha U)}\frac{d\Psi}{d\theta},\label{P}
\end{eqnarray}
\begin{eqnarray}
\frac{d\Omega}{d\theta}=-\frac{2\Omega}{\Psi}\frac{d\Psi}{d\theta}.\label{Omega2}
\end{eqnarray}
where $\beta_\circ=\frac{P_\circ}{B^2_\circ/8\pi}$. Integrating the
last equation, and doing some simplifications we have $\Omega
\Psi^2=K$, where $K$ is an arbitrary constant. Also we can obtain
$\rho$ as follows,

\begin{displaymath}
\rho=-\frac{\beta_\circ\eta_\circ C_{\rm 1}}{2}\left(\frac{8\pi
P\sin\theta}{3U\Psi}\right)^2\left[\frac{1.5(5-3\Gamma)U}{3f(\Gamma-1)}+3\alpha
U^2\right.
\end{displaymath}
\begin{equation}
\left.+\alpha(\frac{dU}{d\theta})^2+\frac{9}{4}\alpha\Omega^2\sin^2\theta+\alpha\sin^2\theta
(\frac{d\Omega}{d\theta})^2\right].
\end{equation}

From equation (\ref{Omega2}) we can find:
\begin{equation}
\frac{d^2
\Psi}{d\theta^2}=\frac{3}{4}\frac{\Psi}{\Omega^2}(\frac{d\Omega}{d\theta})^2-\frac{\Psi}{2\Omega}(\frac{d^2\Omega}
{d\theta})^2 \\
\label{const}
\end{equation}
Comparing equations (\ref{Psi}) and (\ref{const}) we can eliminate
$\Psi$ and then the result can be compared with equation
(\ref{Omega1}), finally we have:
\begin{equation}
3(\frac{d\Omega}{d\theta})^2+2A\frac{d\Omega}{d\theta}-B=0,\label{OO}
\end{equation}
where
\begin{displaymath}
A=\frac{\Omega}{P}\frac{dP}{d\theta}+4\Omega\cot\theta,
\end{displaymath}
\begin{displaymath}
B=\left[\frac{9}{4}+(\frac{1}{\alpha}+\frac{3}{\eta_0})\frac{\rho
U}{c_1 P}\right]\Omega^2,
\end{displaymath}
Equation (\ref{OO}) is an ordinary first order differential equation
for $\Omega^{'}=\frac{d\Omega}{d\theta}$, and it has two roots:

\begin{equation}
\Omega^{'}=\frac{-A\pm \sqrt{{A^2+3B}}}{3}\label{Omega3}
\end{equation}

{substituting the above solution in the main equations of the
system, (\ref{U})-(\ref{Psi})-(\ref{P}), and eliminating $\Psi$  we
obtain:
\begin{eqnarray}
\frac{dP}{d\theta}=&&\frac{3\alpha P}{2(1-\alpha
U)}\frac{dU}{d\theta}+\frac{3\rho KU}{16\pi^2 \beta_0 \eta_0 C_1 P
\Omega^{2} \sin^{2}\theta (1-\alpha U)}\frac{d\Omega}{d\theta}
+\frac{\rho \Omega^2 \sin\theta \cos \theta}{C_1(1-\alpha
U)}\label{P1}
\end{eqnarray}
\begin{eqnarray}
\frac{d^2 U}{d^2\theta}=&&-\frac{2.5}{\alpha}+U-\cot\theta
\frac{dU}{d\theta}-\frac{1}{P}\frac{dP}{d\theta}\frac{dU}{d\theta}
+\frac{\rho}{C_1 \alpha P}(1-\frac{U^2}{2}-\Omega^2
\sin^2\theta)\nonumber\\
&&+\frac{2UK\rho}{\beta_0 \eta_0 \alpha C_1 \Omega }(\frac{3}{8\pi
\rho \sin\theta})\label{U2}
\end{eqnarray}
\begin{eqnarray}
\rho=&&-\frac{\beta_0 \eta_0 C_1}{2}(\frac{8\pi P \sin\theta
\Omega^{\frac{1}{2}}}{3UK^{\frac{1}{2}}})^{2}\left[\frac{1.5(5-3\Gamma)}
{3f(\Gamma-1)}+3\alpha U^{2} +\alpha
\frac{dU}{d\theta}+\frac{9}{4}\alpha \Omega^2
\sin^{2}\theta\right.\nonumber\\
&&\left.+\alpha \sin^{2}\theta
(\frac{d\Omega}{d\theta})^{2}\right]\label{rho1}
\end{eqnarray}

Equations (\ref{Omega3})-(\ref{rho1}) constitute a system of
ordinary non-linear differential equations for the four self-similar
variables, $\Omega, P, U, \rho$. Indeed, the behavior of the
solutions depends on boundary conditions which are supposed based on
some physical assumptions such as symmetry with respect to the
equatorial plane, etc.

There are many techniques for solving these nonlinear equations.
Analytical methods can yield solutions for some simplified
problems. But, in general this approach is too restrictive and we
have to use the numerical methods. Here, one can employ the
method of relaxation to the fluid equations (Press et al. 1992).
In this method we replace ordinary differential equations by
approximate finite-difference equations on a grid of points that
spans the domain of interest. The relaxation method determines
the solution by starting with a guess and improving it,
iteratively. Based on it, this system of equations can be solved
for all unknowns as a function of $\theta$, once if we have a
complete set of boundary conditions which put some physical
constraints on the flow. The boundary conditions are distributed
between the equatorial plane, $\theta=\frac{\pi}{2}$ and the
rotation axis, $\theta=0$. The boundary conditions at $\theta=0$
are:
\begin{equation}
\frac{dU}{d\theta}=\frac{d\Omega}{d\theta}=\frac{dP}{d\theta}=\frac{d\Psi}{d\theta}=
0~~~,~~~U=0,\label{condi1}
\end{equation}
and in this method the boundary conditions at
$\theta=\frac{\pi}{2}$ are}:
\begin{equation}
\frac{dU}{d\theta}=\frac{d\Omega}{d\theta}=\frac{dP}{d\theta}=\frac{d\Psi}{d\theta}=
0.\label{condi2},
\end{equation}

We have found condition $\frac{d\Psi}{d\theta}=0$ by the relation
$\Omega \Psi^2=K$ for both two boundaries which demand the
magnetic flux enclose by the polar axis goes to zero. So field
line thread the equator vertically. We have a limitation to reach
$\theta=0$ numerically. So we try to use boundary condition very
close to polar axis with very small value of $\theta$. We can
obtain physical solutions if we consider the big value of
$\Omega$ and so $\Psi\simeq0$ at $\theta=\epsilon$. Also, at
$\theta=\frac{\pi}{2}$ the equations (\ref{Omega3}) and
(\ref{rho1}) gives:
\begin{displaymath}
\rho=\frac{P}{U}\delta,
\end{displaymath}
\begin{displaymath}
U\delta=\gamma P\left[\frac{3}{2}\epsilon^{'}+3\alpha
U^2+\frac{9}{4}\alpha\right],
\end{displaymath}
where
\begin{displaymath}
\epsilon^{'}=\frac{3(5-\Gamma)}{2f(\Gamma-1)},
~~~~~\delta=-\frac{9\alpha C_1 \eta_0}{4(\eta_0+3\alpha)},
~~~\gamma=-\frac{32\beta_0 \eta_0 C_1 \pi^2}{9K},
\end{displaymath}
The boundary conditions on the above equations require that
variables are assumed to be regular at the endpoints. Also the net
mass accretion rate (\ref{M}) provides one boundary condition for
$\rho$:
\begin{displaymath}
\int_{0}^{\frac{\pi}{2}}\rho(\theta)U(\theta)\sin\theta
d\theta=-\frac{1}{2},
\end{displaymath}
For solving the MHD equations numerically, we need these boundary
conditions. Using these boundary conditions require to choose $K,
\epsilon^{'}, \delta, \gamma$ properly. So we are going to have a
astrophysical approximations for this variables.

For our illustrative parameters, we consider approximate
equipartition of magnetic and kinetic energies for matter flowing
into a neutron star. In quasi-spherical accretion flows,
approximate equipartition proposed by Shwartsman (1971),
\begin{equation}
\frac{B^2}{8\pi}\simeq\frac{1}{2}\rho u_r^2,
\end{equation}
where $u_r$ corresponds to an Alf\'ven speed
$u_A=\frac{B}{\sqrt{4\pi\rho}}\simeq u_{\rm {ff}}$ (free-fall
speed). The density of the infalling materials can be reasonably
approximated by free-fall behavior. In the accretion disks around
the magnetized compact object, due to the complicated nonlinear
effects from the interaction between accretion materials and the
magnetic fields of the compact object, the mechanisms that led to
the magnetic field chandeled accretion flow still remain unsolved.
However, it is widely accepted that due to strong gravity of the
compact object, a very steep and supersonic flow hits the
magnetosphere boundary and deforms its structure drastically. The
infalling flow is halted at distance $r\sim r_A$ where the magnetic
energy density balances the flow kinetic energy, or $u_A=u_{\rm
{ff}}$. Here $u_A$ is calculated from free fall density and
velocity:
\begin{equation}
\rho_{\rm {ff}}=\frac{\dot{M}}{4\pi r_{\rm A}^2 u_{\rm ff}},
~~~~~~~~~~~u_{\rm {ff}}=\sqrt{\frac{2GM}{r}}
\end{equation}
where $\dot{M}$ is the mass accretion rate and $r_A$ Alf\'ven
radius. Interestingly, the numerical simulations (Paatz $\&$
Camenzind 1996, Koldoba, Lovlace $\&$ Ustyugova 2002) show that at
this distance $r\sim r_A$ the material funnels through the
magnetosphere and fall onto the magnetic poles. The funnel flow is
found to reach the star's magnetic poles with the velocity close to
that of free-fall velocity.

The magnetic configuration outside of a compact star, such a neutron
star, is a dipole field then in spherical polar coordinates we have
\begin{equation}
\textbf{B}=B_pR^3\left(\frac{2\cos\theta}{r^3}\textbf{e}_r
+\frac{\sin\theta}{r^3}\textbf{e}_\theta\right)\label{B}
\end{equation}
where R is the radius of the neutron star and $B_p$ is the magnetic
field at the magnetic poles. Thus in the equatorial plane
($\theta=\frac{\pi}{2}$), the equation (\ref{B}) becomes
$B\simeq\frac{B_pR^3}{r^3}$. So we obtain,
\begin{equation}
\frac{\dot{M}}{4\pi r_{\rm A}^2 u_{ff}}\simeq\frac{B^2_pR^6}{8\pi
GMr_{\rm A}^5}\label{rho}.
\end{equation}
The equation (\ref{rho}) implies that the infalling mass close to
the equatorial plane of a neutron star is stopped at a radial
distance,
\begin{equation}
r_{\rm A}=B_p^{4/7}R^{12/7}{\dot{M}}^{-2/7}(2GM)^{-1/7}\label{r}
\end{equation}
for $M\sim M_\odot$, ${\dot{M}}\sim 1.85\times 10^{16} g s^{-1}
(\dot{M}<\dot M_{Edd}=1.39\times 10^{18}\frac{M}{M\odot} g s^{-1})$,
$B_p\sim10^{12} G$ and $R\sim10^6cm$, the equation (\ref{r}) gives
$r_\circ\sim r_{\rm A}\sim 5.2\times 10^8 cm$. Thus we can obtain
$B_\circ=B_p\left(\frac{R}{r_\circ}\right)^3\sim10^6 G$ and
$\rho_\circ\sim3.6\times 10^{-10} g ~{cm}^{-3}$. On the other hand,
the gas pressure, density and temperature are related by
\begin{equation}
p_\circ=\frac{\rho_\circ kT}{\mu_\circ m_H}
\end{equation}
where $\mu_\circ$ is the mean molecular weight, k Boltzman constant
and $m_H\sim m_p$ is the mass of the hydrogen atom. We assume the
accreting materials to be mainly hydrogen and to be nearly ionized.
For accreted materials, the potential energy released is
$\frac{GM_*M_{acc}}{r_\circ}$ and the thermal energy is
$2\times\frac{3}{2}kT$ ($T_{\rm vir}=\frac{T_{\rm th}}{2}$);
therefore, $T_{th}=\frac{GM_*M_{\rm
acc}}{3kr_\circ}\sim5.5\times10^9K$ we obtain $p_\circ\sim3.2\times
10^8 dyne~cm^{-2}$. Consequently, in turn it can be reduced that
$\beta_\circ=0.01$, $ C_1=0.8$ and $ C_2=2\times10^3$.

Imposing two boundary conditions (\ref{condi1}) and (\ref{condi2})
to the equations (\ref{Omega3})-(\ref{rho1}), we obtain numerical
solutions for the flows with a fixed value of $\eta_\circ=0.1$,
$\Gamma=4/3$ and $\alpha=0.01, 0.05$ and $0.1$ with a sequence of
the increasing values of parameter f or decreasing cooling. Figures
1-5 show our results.
\input{epsf}
\epsfxsize=4.5in \epsfysize=3.5in
\begin{figure}
\centerline{\epsffile{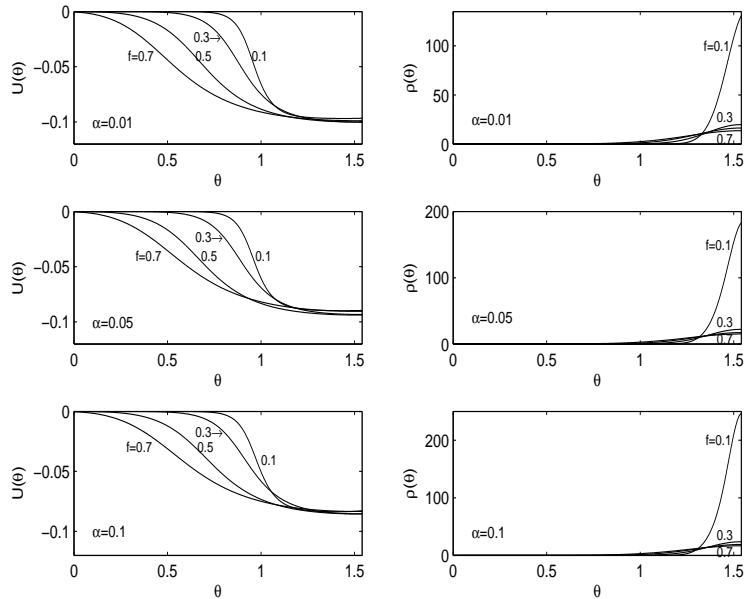}}
 \caption{The self-similar solutions of radial velocity $U$ (left panels) and density $\rho$ (right panels)
      as a function of polar angel $\theta$ corresponding to
      for Top: $\alpha$=0.01, Middle: $\alpha$=0.05 and Bottom: $\alpha$=0.1.}
 \label{figure1}
\end{figure}
\section{Results: Typical Solutions}
\baselineskip17pt \textheight 23cm
We have obtained numerical solutions of equations
(\ref{Omega3})-(\ref{rho1}) for variety of values of the viscosity
parameter $\alpha$ and advection parameter $f$. Figures 1-4 show a
typical sequence of solutions correspond to $\alpha=0.01,0.05,0.1$
and $f=0.1, 0.3, 0.5, 0.7$. The solutions may be considered either
as flows with a fixed value of viscosity parameter and with a
sequence of increasing advection parameter or decreasing cooling
(Figures 1 and 3) or a sequence of different values of viscosity
parameter $\alpha$ in a fixed advection regime (Figures 2 and 4).

The six panels in Figure 1 show the variations with respect to the
polar angle $\theta$ of various dynamical quantities in the
solutions. The top left panel displays the dimensionless radial
velocity $U(\theta)$ as a function of $\theta$. The velocity is zero
at $\theta=0$ (this is a boundary condition) and maximum at
$\theta=\frac{\pi}{2}$. So the maximum accretion velocity is at
equatorial region and on the polar axis there is no mass inflow. As
we expected the velocity is sub-Keplerian. We find that in the
boundary, $U(\theta)$ is essentially independent of advection
parameter $f$. But in the intermediate, the radial velocity is
modified by $f$; in the SH2004 solutions, two distinct regions in
the $U(\theta)$ profile could be recognized. The bulk of accretion
occurs from equatorial plane at $\theta=\frac{\pi}{2}$ to a surface
at $\theta=\theta_s$, inside of which the radial velocity is zero.
While NY1994 solutions there is no zero inflow in
$0<\theta<\frac{\pi}{2}$. Our solutions show that in a given
$\theta$ the radial velocity is increased when we increase the
advection parameter. The middle and bottom left panels display the
radial velocity in $\alpha= 0.05, 0.1$ for a sequence of advection
parameters respectively.

The top right panel shows profile of the density $\rho(\theta)$. The
density contrast in the equatorial and polar regions increase with
decreasing advection parameter $f$. The density grows and becomes
concentrated toward the equatorial plane. For a given $\alpha$,
solutions with small values of $f$ behave like standard thin disks,
as might be expected since these solutions correspond to
$f\rightarrow 0$ and so advect very little energy. In the opposite
advection-dominated limit, which corresponds to $f\rightarrow 1$,
our solutions describe nearly spherical flows which rotate at much
below the Keplerian velocity. This is demonstrated in Figure 5 where
we display iso-density contours in the meridional plane. The middle
and bottom right panels display the density profiles in $\alpha=
0.05, 0.1$ for a sequence of advection parameters respectively. This
advection-dominated solutions have very similar properties to the
approximated solutions derived by NY1994 and SH2004. The results
show that these quantities are very sensitive to the advection
parameter. For low advection regimes, $f=0.1, 0.3$, we have to
separate the regime for radial inflow. But for high advection
regimes $f=0.5, 0.7$, the radial velocity is non zero around the
pole.
\input{epsf}
\epsfxsize=4.5in \epsfysize=3.5in
\begin{figure}
\centerline{\epsffile{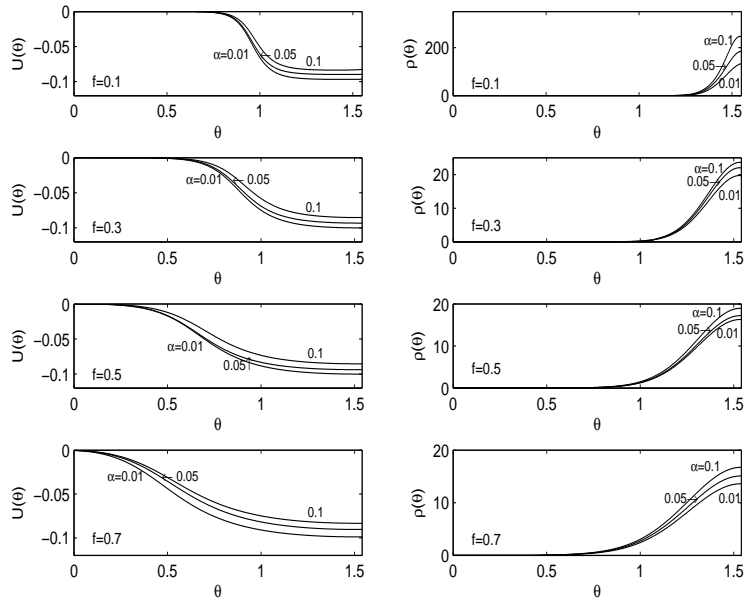}}
 \caption{ The self-similar solutions of radial velocity $U$ (left panels) and density $\rho$ (right panels)
 as a function of polar angel $\theta$ corresponding to
      $\Gamma=4/3$, $\eta_{0}=0.1$, $\beta_0=0.01$ and $\alpha=0.01, 0.05, 0.1$ for
      $f=0.1, 0.3, 0.5, 0.7$ from Top to Bottom, respectively. }
 \label{figure2}
\end{figure}
\baselineskip20pt \textheight 21cm
The behavior of the solutions, self-similar radial velocity and
density profile, are demonstrated in Figure 2 for different values
of advection parameters with variation of viscosity parameters. The
panels in Figure 2 show that For a fixed $f$, the density maximum
increases with increasing viscosity parameter while the radial
velocity decreases with increasing it. In this case, increasing the
viscosity parameter corresponds to increasing heating mechanisms, so
in a fixed advection regime, there are more energy to advect into
central star. The top Panels in Figure 5 Show that the disk to be
thick. The solution with the same $f$ but with different values of
$\alpha$ are virtually indistinguishable from one another. There are
probably, more significantly variations when $\alpha$ exceeds above
$0.1$. Recently King et al. 2007 assart that in a thin and fully
ionized disk the best observational evidence suggest a typical range
$\alpha\sim 0.1-0.4$ where relevant numerical simulations tend to
drive estimates for $\alpha$ which are an order of magnitude
smaller. However, such large values of $\alpha$ are probably
unlikely (eg. Narayan, Loab, $\&$ Kumar 1994; Hawley, Gammie, $\&$
Balbus 1994), and so we have not explored this region of the
$\alpha$ parameter.
\input{epsf}
\epsfxsize=4.5in \epsfysize=3.5in
\begin{figure}
\centerline{\epsffile{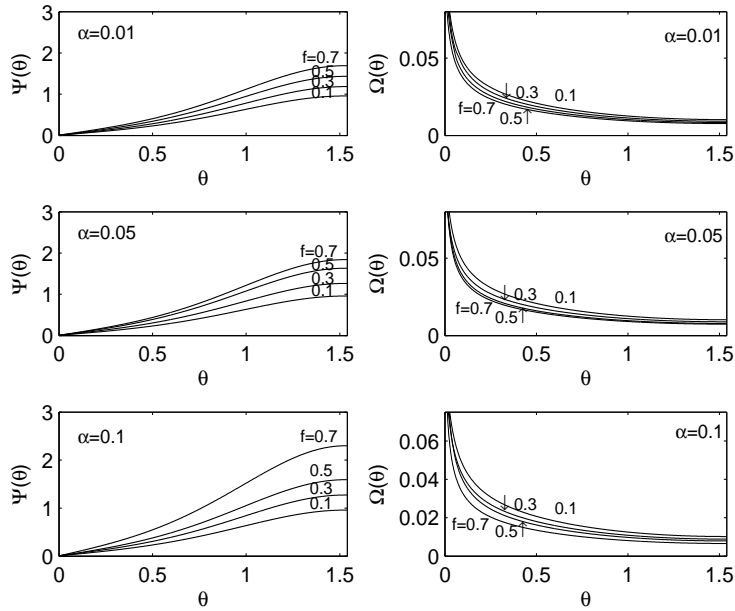}}
 \caption{The self-similar solutions of magnetic flux $\Psi$ (left panels) and angular momentum $\Omega$ (right panels) as a function
          of polar angel $\theta$ corresponding to $\Gamma=4/3$, $\eta_{0}=0.1$, $\beta_0=0.01$, $f=0.1, 0.3, 0.5, 0.7$
          for Top: $\alpha$=0.01, Middle: $\alpha$=0.05 and Bottom: $\alpha$=0.1 in rotating accretion flows.}
 \label{figure3}
\end{figure}
The $U(\theta)$ and $\rho(\theta)$ profiles both peak at
$\theta=\frac{\pi}{2}$. Therefore, in all our solutions the bulk of
accretion occurs along the equatorial plane and accretion rate goes
to zero along the rotational pole. An interesting feature of these
solutions, as already mentioned, for low advection ,$f$, we have a
thin disk; in all cases there is a low density with a higher
temperature corona above the disk.

Figure 3 displays the magnetic flux function and the angular
velocity for different values of advection parameter in a fixed
viscosity parameter. Integration of equation (\ref{Omega2}) yields
$\Omega \Psi^2=K$, where $K$ is an arbitrary constant. So angular
velocity and magnetic flux function should have opposite behavior
along $\theta$ direction. We plot them with choosing proper input
parameters which introduced in the end of the last section. Figure 3
shows that magnetic flux function, $\Psi(\theta)$, varies by only
$\sim 50$ percent. The magnetic flux function increases by
increasing advection in accretion disk in a fixed viscosity. The
behavior of angular velocity is exactly opposite. Figure 4 displays
the behavior of the magnetic flux and angular velocity for different
values of viscosity parameter in a fixed advection. The solutions
implies that in a fixed, low $f$, the effect of different $\alpha$
is indistinguishable.
\input{epsf}
\epsfxsize=4.5in \epsfysize=4.5in
\begin{figure*}
\centerline{\epsffile{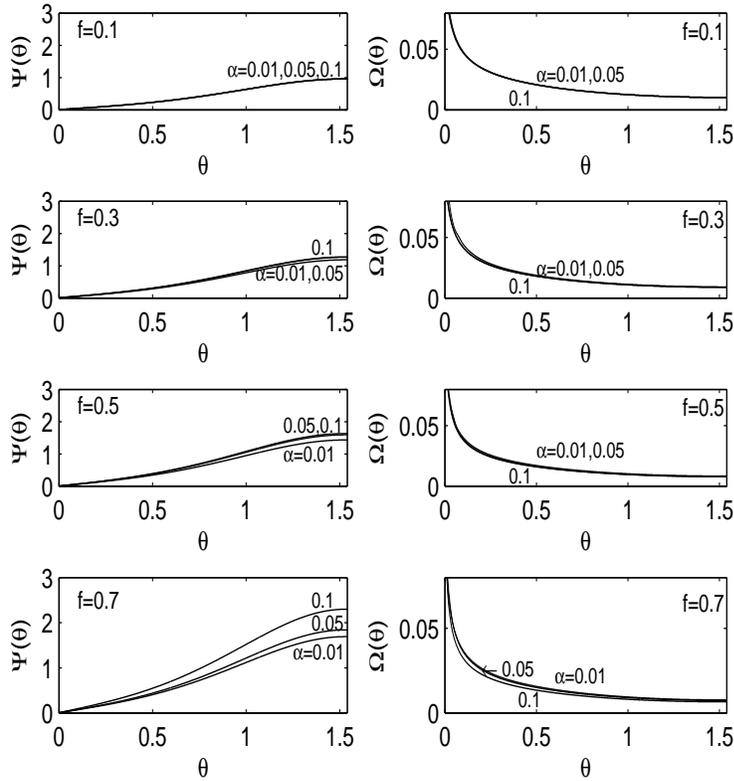}}
 %\vspace{4mm}
 \caption{The self-similar solutions of magnetic flux $\Psi$ and angular momentum $\Omega$ as a function of polar
      angel $\theta$ corresponding to
      $\Gamma=4/3$, $\eta_{0}=0.1$, $\beta_0=0.01$ and $\alpha=0.01, 0.05, 0.1$ for
      $f=0.1, 0.3 ,0.5, 0.7$ from Top to Bottom, respectively.}
 \label{fig4}
\end{figure*}
\input{epsf}
\epsfxsize=3.5in \epsfysize=3.5in
\begin{figure*}
\centerline{\epsffile{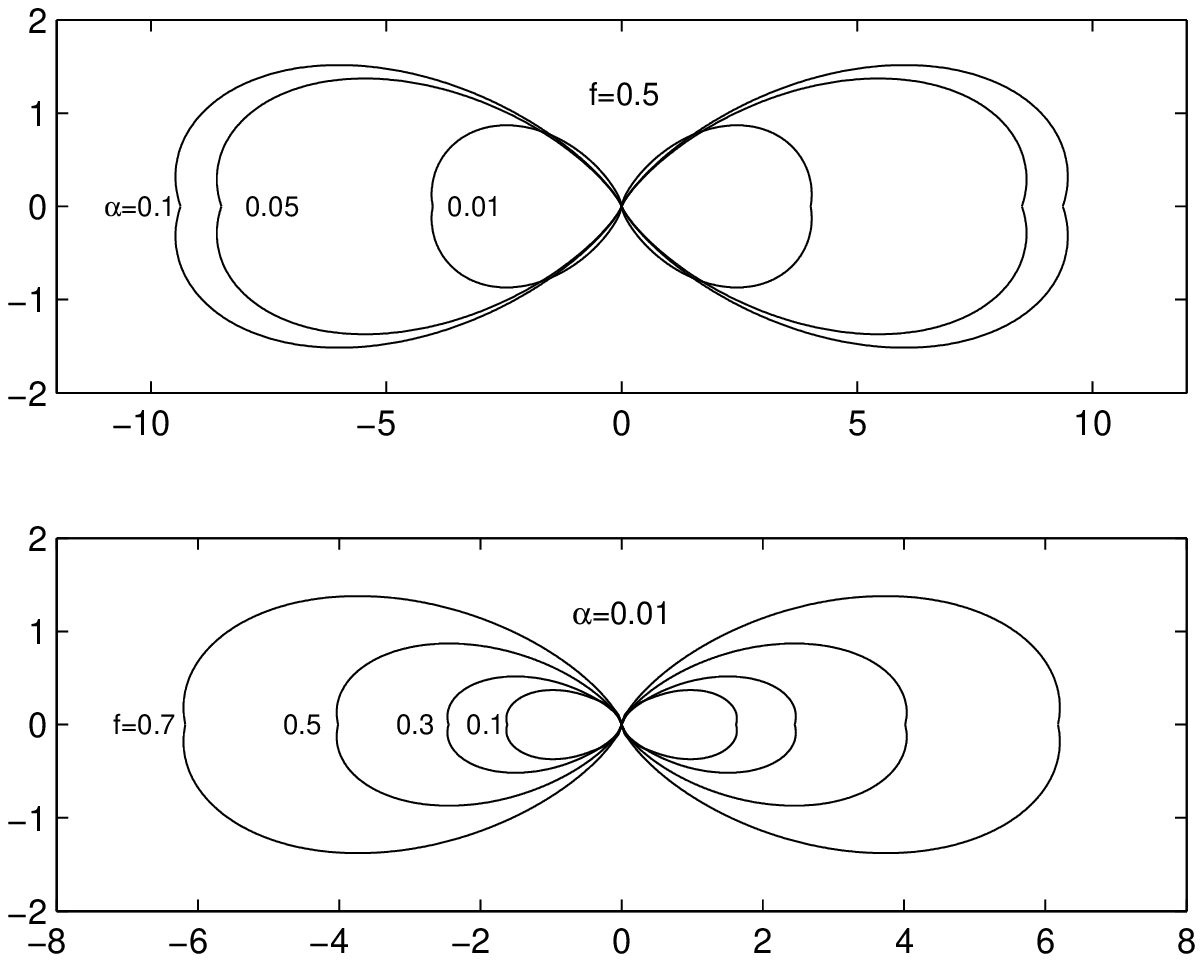}}
 %\vspace{4mm}
 \caption{Isodensity contours in the meridional plane for  $\Gamma=4/3$, $\eta_{0}=0.1$,
      $\beta_0=0.01$ for Top: $f$=0.5 and $\alpha$=0.01, 0.05, 0.1, Bottom: $\alpha$=0.01
      and $f$=0.1, 0.3, 0.5, 0.7.}
 \label{fig5}
\end{figure*}
\section{SUMMERY AND CONCLUSION }

The main aim of this investigation was to obtain axisymmetric
self-similar advection-dominated solution for viscose-resistive
accretion flow. We have presented the results of self-similar
solutions of the effect of the viscosity and rotation on
magnetically driven accretion flows from a flow threaded by poloidal
magnetic fields where only serious approximation we have made is the
use of an isotropic $\alpha$ viscosity and constant magnetic
diffusivity. Attention has restricted to flow accretions in which
self-gravitation is negligible. We included the magnetic diffusivity
so that its value was constant throughout our analysis. Using the
basic equations of fluid dynamics in spherical polar coordinates
$(r,\theta,\phi)$, we have employed the method of self-similar for
thick discs to derive a set of coupled differential equations which
govern the dynamics of the system. We then solved the equations by
the method of relaxation by considering boundary conditions and
using $\alpha$-prescription (Shakura \& Sunyaev, 1973) in order to
extract some of the similarity functions in terms of the polar angle
$\theta$. Figures are considered for $\alpha=0.01, 0.05,$ and $0.1$
so that for any $\alpha$ we considered $f=0.1, 0.3, 0.5$ and $0.7$.

We showed that the radial and rotational velocities are well below
the Keplerian velocity. The Bulk of accretion with nearly constant
velocity occur in the region which extend from equatorial plane to a
given $\theta$ which highly depends on advection parameter $f$. In a
non-advective regime, low $f$, we have a standard thin accretion
disk but for high $f$ the accretion is nearly spherical. The
geometrical shape of the flow is determined by the amount of
viscosity and advection, in a fixed magnetic diffusivity. The
accretion disk with efficient cooling ($f\rightarrow 0$) has
low-density, high temperature corona which implies that the regular
thin disks may be accompanied by advection-dominated corona which
can drive low-density wind.

Our results show the flow of ionized accretion materials is not
disklike in morphology. The closest along our solution in the
accretion literature is Bondi (1952) spherical accretion. Our flows
differ in important ways from Bondi problem. The gas in our model
rotates and has viscose interaction through which angular momentum
is transported outward. Also, our flow has magnetic interaction with
dipole magnetic field of the central star which it can redistribute
the angular momentum within the accretion disk. The angular velocity
is significantly sub-Keplerian, and this may have an important role
in spin-up of accreting stars. Stars which their spins grow by
interacting with accretion materials are likely to reach a steady
state with a rotation rate below the break-up limit. This
advection-dominated solution may be a good solution for angular
momentum problem in the star formation.

However, our results improve the physics of advection-dominated
accretion disks. It is important that the magnetic diffusivity can
modify the dynamical quantities of the disks. We developed NY1994
solutions to a realistic model for ADAFs by adding magnetic
diffusivity. But in future we are going to investigate the effect of
non-constant magnetic diffusivity, $\eta_0$. Several developments
can be investigated to reach a much realistic description of the
physics of accretion disks around the magnetized compact objects.

\end{document}